\documentclass[11pt]{article}
\usepackage{graphicx}
\usepackage{hyperref}      
\usepackage[margin=0.75in]{geometry}

\title{3D Technologies for Large Area Trackers}

\begin{document}

\def\bibname{References}
\bibliographystyle{plain}

\raggedbottom

\pagenumbering{roman}

\parindent=0pt
\parskip=8pt


\pagenumbering{arabic}


\maketitle

\begin{center}\begin{boldmath}

Whitepaper Submitted to Snowmass 2013\\

 \end{boldmath}
\end{center}
G. Deptuch$^a$,
U. Heintz$^b$,
M. Johnson$^a$, 
C. Kenney$^d$,
R. Lipton$^a$,
M. Narian$^b$,
S. Parker$^e$,
A. Shenai$^a$, 
L. Spiegel$^a$, 
J. Thom$^c$,~ and Z. Ye$^f$\\
\llap{$^a$}Fermilab,
  P.O. Box 500, Batavia, Il , USA,
\llap{$^b$}Brown University,
  Providence, RI , USA,
\llap{$^c$}Cornell University,
  Ithaca, NY , USA,
 \llap{$^d$} SLAC National Accelerator Laboratory,
Menlo Park, CA, USA,
 \llap{$^e$} University of Hawaii,
 Honolulu, HI, USA 
  \llap{$^f$} University of Ilinois, Chicago,
 Chicago, Il, USA 
\\

\bf{Abstract}\\
\rm
We describe technologies which can be developed to produce large area, low cost pixelated tracking 
detectors.  These utilize wafer-scale 3D electronics and sensor technologies currently being developed in industry. 
This can result in fully active sensor/readout chip tiles which can be assembled into 
large area arrays with good yield and minimal dead area.  The ability to connect though the bulk of the 
device can also provide better electrical performance and lower mass.



\section{Introduction}
In the next generation of collider experiments  detectors will be challenged by unprecedented luminosity and 
data volumes. We will need to build large area arrays (100's of meter$^2$) of highly pixelated detectors with minimal 
dead area and at reasonable cost.  Current fine pitch bump bonding technologies require individual placement 
of chips on sensors followed by a solder melt cycle. Technologies which provide wafer-scale interconnect offer 
prospects of lower cost, finer pitch and lower mass interconnects between sensors and readout chips. In addition 
3D interconnect technology allows connections through the bulk of the silicon, providing low inductance and 
capacitance paths for signal and power connections.
Table~\ref{costs}  summarizes current and projected costs and yields of various bonding
technologies. 

In the past pixel sensor arrays have been limited in the module area by:
\begin{enumerate}
\item Space needed at the edges of the detectors to reduce the field near the damaged dicing cut regions.  These regions act 
as charge emitters and can cause unacceptably large currents in the edge strips.

\item Die size of the Readout Integrated Circuit (ROIC), determined by the die yield and reticule area

\item The need to provide edge locations for wire bonds for connection to the ROIC.
\end{enumerate}
The development of active edge sensors by C. Kenney, S. Parker and collaborators has addressed the first problem~\cite{Kenney:2006ky}~\cite{Kenney2001}
\cite{Cinzia2009}.  Recent work 
on the large area FEI4 chip~\cite{FEI4} has shown that a large pixel chip can be produced with good yield.  Processes related 
to 3D electronics can solve the third problem. The combination of all three can result in a wafer scale fabrication 
of tiles or arrays which can be integrated into large modules with high yield and relatively low cost. 
Using such tiles, large area pixelated modules can be assembled with known good integrated sensor/readout die with  
 large pitch backside bump bond interconnects.  
 
 \begin{table}[htdp]
\caption{Current and projected costs and yields for sensor/readout integration technologies.}
\begin{center}
\begin{tabular}{|c|c|c|c|}
\hline
Component & Current or projected cost & Yield & Comment \\
\hline
Readout IC & \$8/cm$^2 \cite{3DCost} $ & 65-70\% \cite{FEI4Yield} & Current 3D wafer cost and  \\
& & & FEI4 prototype yield \\
\hline
Active Edge Sensors & \$53/cm$^2$ &  $\approx$90\% & Current cost for prototype \\
& & & 150 mm wafers \\
\hline
Silicon Strip Sensors &  \$10/cm$^2$ & $\approx$100\% & CMS tracker costs \\
\hline
Bump bonding & \$213/cm$^2$ & 98\%\cite{merkel} & CMS forward pixel costs\\
(2007) & & & Yield $\equiv < $20 bad bumps/chip \\
\hline
Bump bonding & \$62/cm$^2$ & - & CMS forward pixel upgrade\\
(2012) & & & \\
\hline
Wafer scale & \$0.04/cm$^2$ & $\approx$90\% & Projected by Yole Development \cite{DBICost}\\
DBI bonding  & & & for high volume production \\
\hline
Target Costs (2020s) & \$10/cm$^2$ & 80\%  & Assuming 200 mm sensor wafers \\
 & & & and batch active edge process \\
\hline
\end{tabular}
\end{center}
\label{costs}
\end{table}%

\section{Technologies}
\subsection{3D Circuits}
3D circuitry (3DIC) is the generic term for a set of technologies, including wafer bonding, thinning, and interconnect, which 
allows vertical interconnection of multiple layers of CMOS electronics~\cite{ibm3d}~\cite{3Dint}. 
3D interconnects have the advantage of reducing inductance and capacitance while increasing circuit density and allowing 
the integration of heterogeneous device types. The key enabling technology is the Through-Silicon-Via (TSV),
a metal-filled hole etched into the wafer bulk silicon which forms the conducting path between tiers of a multi-layer assembly.
 In high energy physics 3D circuitry would allow us to directly integrate 
sensors and their readout electronics without the use of expensive and cumbersome fine pitch bump bonds.

We have explored three technologies for 3D devices.  Our initial studies were with MIT-Lincoln Labs and used their 0.18 
micron three-tier process.  This process utilizes oxide bonding to join the tiers and vias are etched through the SOI buried 
oxide after the tiers are bonded.  
We developed a demonstration 3-tier ILC vertex chip (VIP) in the MIT-LL process~\cite{Deptuch:2011zz}.  
We have used the Ziptronix oxide bonding process with imbedded metal to mate BTeV FPiX ROIC wafers to sensors fabricated at 
MIT-Lincoln Labs~\cite{ye}.  Fermilab has also sponsored a two-tier 0.13 micron CMOS 3D IC run with 
Tezzaron/Global Foundries~\cite{Deptuch:2010zza}~\cite{YaremaFEE11} that features $1.2~\mu$ diameter, $6\mu$ deep tungsten filled TSVs.  
In this process wafers were bonded face-to-face utilizing either copper thermocompression or the Ziptronix 
Direct Bond Interconnect (DBI) oxide bonding process described below.  Bond pitch for these wafers was 4 microns. 
After bonding the top wafer is thinned to expose the tungsten through-silicon-vias and the top 
is patterned to provide contacts for bonding. The final set of wafers from Tezzaron have  been received. 
The copper-copper wafers suffered from alignment problems, lowering the yield of good chips, 
but the DBI wafers have both good alignment and yield.

\begin{figure*}[ht]
\centering
\includegraphics[width=150mm]{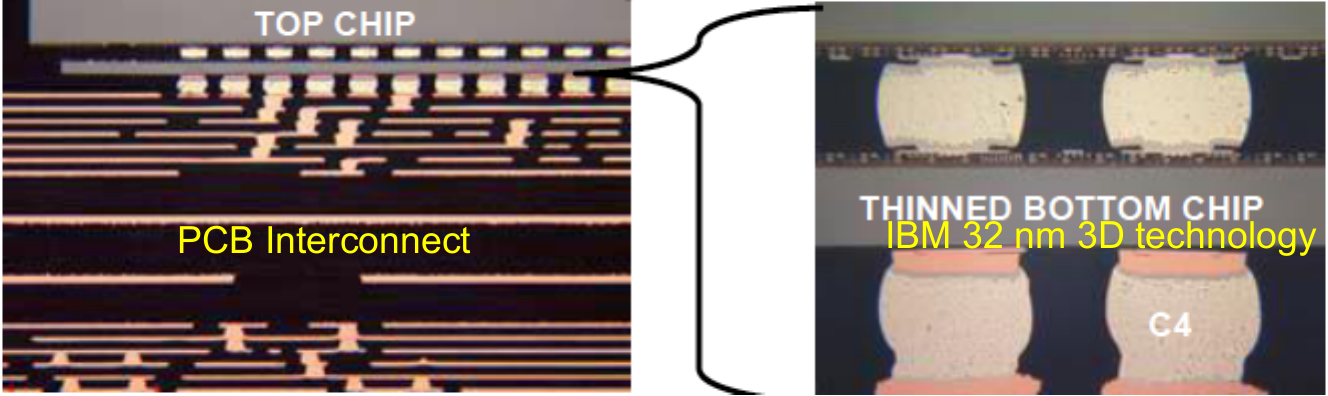}
\caption{Cross section of a 3D assembly utilizing the IBM 32 nm copper stud 3D interconnect process.} \label{IBM_3D}
\end{figure*}

 Other 3D processes are becoming commercially available.  IBM and Micron Semiconductor will soon announce a 3D stacked 
 memory product based on IBM's 32 nm TSV process~\cite{shapiro} with copper stud bonding shown in figure~\ref{IBM_3D}. 
 This technology is based on 
 die-to-die bonding, potentially maximizing yield by utilizing known good die from both sensors and ROICS.  
 Combining this process with active edge sensors 
 could provide a very appealing path toward active tiles which could be used to build large area sensors.  CEA-LETI is offering 
 "Open 3D" services for prototyping and/or low volume production.  Via-last processes, such as provided by LETI, 
 can provide TSVs in processed wafers, with the limitation that the wafers must be thick enough to handle (>100 microns) 
 and the TSV aspect ratio is generally limited to less than 20:1.

\subsection{Active Edge Sensors}
Active edge sensors are an outgrowth of work done to develop 3D sensors, which provide good charge collection combined 
with radiation hardness.  The technique utilizes a 
deep reactive ion etch of silicon to create a nearly vertical trench with smooth edges. The high quality of the trench wall 
avoids charge generation normally associated with saw-cut edges and allow closer placement of adjacent sensors.
The trenches are filled with doped polycrystalline silicon.  The dopant is diffused into the surrounding single crystal silicon and annealed for activation.  The dopant density gradient will make an electric field in the collection direction.  These steps may be done at the same time for the other like-type electrodes.  The depth of the doped silicon must be great enough so it is not depleted by the largest applied bias voltage.  Mechanical integrity is maintained by bonding the sensor to a support wafer.  The oxide bond also forms an etch stop for the trenching and singulation processes.    In the 
case of an oxide bonded handle, in a silicon-on-insulator structure, the bond also forms an etch stop for subsequent trenching and 
singulation processes. An alternative active edge technology based on 
wafer cleaving and atomic layer deposition~\cite{Slim} has the prospect of achieving similar goals without 
the additional processing needed in the deep trench process.  

\subsection{Oxide Bonding}
Bonding of silicon wafers is a key enabling technology for nanotechnology, micromachining, and 3D electronics. 
 A variety of techniques have been developed, including adhesive bonding, metal eutectic bonding, and bonding 
 based on the silicon oxide surface either grown or deposited on a wafer~\cite{tong}. Oxide bonding has the 
 advantages of being mechanically robust, chemically inert, and capable of withstanding the high temperatures 
 typical of silicon processing.
  
 The direct oxide bond~\cite{tong2} is formed by bringing together silicon wafers which have been planarized and chemically 
 treated to form a hydrophilic surface.  When the wafers are brought together at room temperature a van der Walls bond forms
 between the hydrogen atoms at the wafer surface.  Further annealing above $150^{\circ}{\rm C}$ causes the formation of covalent 
 hydrogen bonds and provides a substantial increase in bond strength.  The Ziptronix DBI process  
 imbeds nickel or copper in the planarized oxide surface.  The metal forms an interconnect  to a seed metal layer 
 in the resulting oxide bonded wafers 
 which can be used to build 3D interconnect structures~\cite{enquist}. The process requires good planarity, 
 which is a feature of modern CMOS processes.  Dust particles present during the bonding 
 process can produce local bond voids.  These unbonded areas limit the large area module yield and are a 
 motivation for active tile development. The DBI process can be used for wafer-to-wafer or 
 chip-to-wafer bonding. However the wafer to wafer process is least expensive and most well suited to 
 scaling to large volumes.

 \subsection{Interconnect}
 Once bonded to a sensor, connections need to be made to the ROIC, which is now face down. 
 Normally in a bump bonded pixel detector assembly the chip area is larger than that of the sensor with bond pads extending out of the edges.  
 In an oxide bonded assembly signals as well as power need to be brought in through the bulk silicon of the ROIC. 
 If the ROIC has imbedded through silicon vias the wafers can be thinned and vias exposed using the same technique
utilized in the Tezzaron two-tier 3D integrated circuit run.  A non-TSV wafer can be used by thinning the top silicon to  $\approx 10\mu$, etching 
 that silicon to the I/O pad contacts, and depositing a redistribution metal layer to provide the final contacts through 
 what normally would be the bottom of the ROIC. The 
 redistribution layer can provide interconnects between adjacent chips, route power, and bump bond pads. 
 This layer can also provide "stitching" between adjacent reticules to provide a large effective area which can be 
 defined at the dicing stage.
 
 \begin{figure*}[ht]
\centering
\includegraphics[width=100mm]{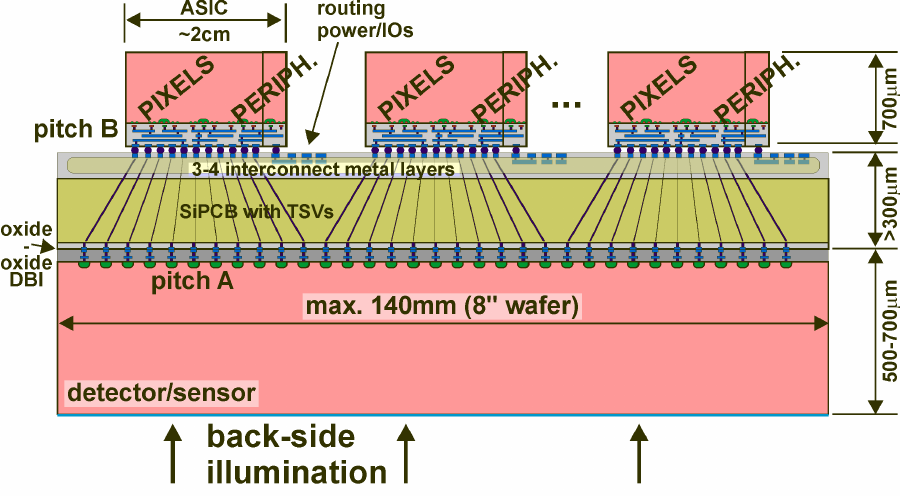}
\caption{Sketch of an interposer-based large area array.} \label{Interposer}
\end{figure*}
 
 An alternate interconnect approach is to utilize so-called 2.5D interconnect. This technology utilizes a 
 silicon or glass interposer which incorporates TSVs and redistribution of signals (figure~\ref{Interposer}).  This approach decouples 
 the 3D processing from the sensors and readout chips and offers a substrate with very fine interconnect 
 capability and coefficient of thermal expansion match to the silicon ROICs and sensors. An 
 interposer-based solution could partially decouple sensor and ROIC pitches, allowing for a variety of 
 external interconnects, and allow for conventional bump bonding of ROICS. 

\section{Large Area Assembly}
We  now consider techniques for the fabrication of large area arrays utilizing the tools 
described in section 2. All are based on wafer-to-wafer bonding of sensors and ROICs 
utilizing the DBI process.  

\begin{figure*}[ht]
\centering
\includegraphics[width=100mm]{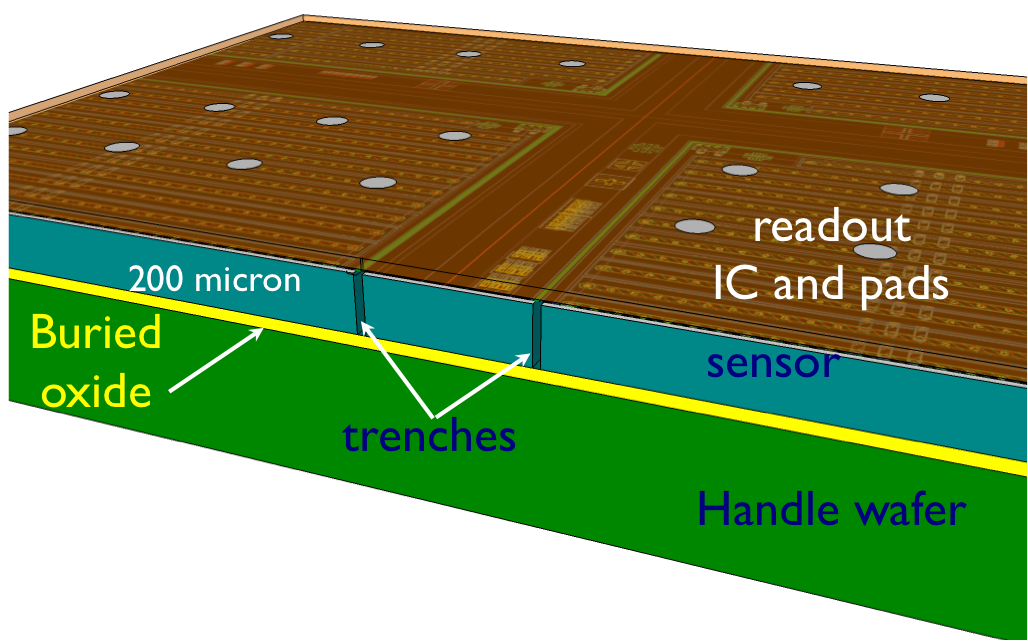}
\includegraphics[width=100mm]{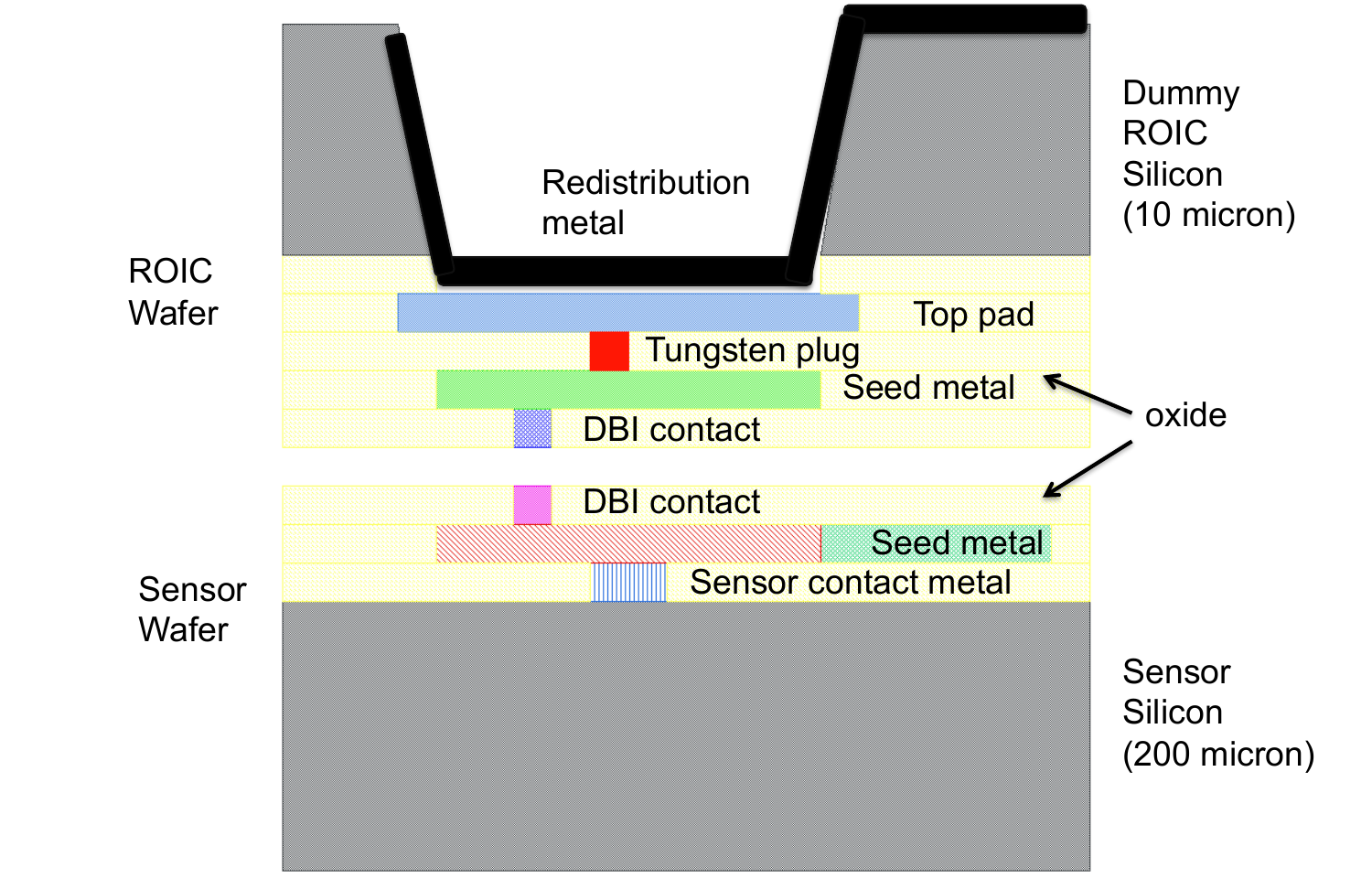}
\caption{Top - 3D sketch showing a view of the wafer stack, including handle wafer, trenches, and 
the region between reticules. Bottom- Schematic view of the final stack with DBI contact layers 
and sensor and top contacts.} \label{SOI_stack}
\end{figure*} 

\subsection{Tiles with Active Edge Sensors}
We are currently exploring a process which utilizes "standard" active edge sensors. These 
are n-on-p devices with 200 micron thick sensors in an SOI stack with a 500 micron thick 
handle wafer.  The tiles are fabricated by DBI bonding the sensor wafer stack to a dummy 
ROIC which rearranges the signal contact pattern.  After bonding contacts on the 
dummy ROIC wafer are exposed by grinding and etching away the dummy wafer silicon 
substrate and first layer of oxide to expose what would normally be the bottom surface of the readout 
pads. This surface is then metalized to provide sites for bump bond placement. The stack is 
shown in figure~\ref{SOI_stack}. Finally the active die must be singulated.  This is 
accomplished by etching the 10 microns of dummy 
wafer silicon and 3 microns of oxide, followed by an etch of the 200 microns of polysilicon filling the 
trench.  The SOI oxide forms an etch stop in this process.  The wafer is then attached to a temporary 
handle wafer and the original SOI handle is ground away, leaving isolated tiles.

\subsection{Post-Processed Tiles}
A simpler variant on the above process would DBI bond ROIC and untrenched sensor wafers.  
The tiles are then defined by etching trenches in the backside of the sensor wafer.  The etching 
process is low temperature and should not harm the ROIC.  However the resulting edges would 
not have the doping that normally defines the side electrode in active edge devices.  
Post-trench doping and annealing would not be possible, since the anneal 
would require higher temperatures than could be 
tolerated by the ROIC.  In this case a process similar to that explored for slim-edge fabrication, 
utilizing atomic layer deposition to activate the edges should provide acceptable performance. 
This process avoids the SOI stack needed for the initial sensor wafer, provides a more planar 
surface for oxide bonding, and avoids the complex singulation and handle wafer removal 
necessary with the pre-processed active edge devices.

This process is very similar to plasma dicing, which utilizes the same deep reactive ion etching 
process to dice wafers~\cite{singulate}, so it should be commercially available for both 
200 mm and 300 mm diameter wafers.

\subsection{Die-to-Die tiles}
An alternative which utilizes the die-to-die capability of an IBM-style process could bond 
active edge sensor tiles to ROICs with imbedded TSVs.  This probably would not require 
ROIC and sensor wafers of equal diameter assuming the copper bond preparation process 
was available for the smaller diameter sensor wafers. Although this process might not have the 
economies of scale of a wafer level process, improved yields by utilizing tested die might 
make this process competitive in cost.

\subsection{Interposer-based Array}
If the density of voids in the DBI bonded assembly is sufficiently small an appealing 
assembly would utilize a silicon interposer DBI bonded to the sensor wafer.  Such an 
assembly could, in principle, utilize the largest sensor that can be inscribed 
in the wafer with good yield, typically about 1 cm inset from the wafer edge. The 
interposers redistributes the signals to ROIC bump bonds on the top of the interposer. 
This works well in a situation where the area of the ROIC is smaller that that of the 
associated pixels. 

\section{Prospects}
The ability to build large areas of pixelated arrays with complex readout electronics will be crucial to future detectors.  
Bonding costs and yield will define the limitations of these detectors.  
We have described a plan to demonstrate pixelated tiles or large area arrays that have the 
prospect of making such large area devices affordable.  To complete this development sensor 
wafers which match the 8" diameter of the ROIC wafers are needed.  Such wafers have been 
demonstrated in the SOI process by both Lapis/OK~\cite{Onuki:2011zz} and American Semiconductor. 
We expect that commercial manufacturers will soon establish 8" high resistivity sensor production. 


 


\end{document}